\documentclass[prl,superscriptaddress,showpacs,preprint]{revtex4-1} 
\usepackage{amsmath,graphicx,amssymb,amsfonts}
\begin{document}

\title{Granular Gas: Vibrating Walls,\\ Two-Peak Distribution and Hydrodynamics}
\author{Yanpei Chen}
\affiliation{Key Laboratory of Soft Matter Physics, Beijing National Laboratory for Condensed Matter Physics, Institute of Physics, Chinese Academy of Sciences, Beijing, 100190, China}
\author{Meiying Hou}
\affiliation{Key Laboratory of Soft Matter Physics, Beijing National Laboratory for Condensed Matter Physics, Institute of Physics, Chinese Academy of Sciences, Beijing, 100190, China}
\author{Yimin Jiang}
\affiliation{Central South University, Changsha, China 410083}
\author{Mario Liu}
\affiliation{Theoretische Physik, Universit\"{a}t T\"{u}bingen, 72076 T\"{u}bingen,
Germany}
\date{\today }

\begin{abstract}
Vibrating walls, used to maintain the temperature in a granular gas, modify the system strongly. Most conspicuously, the usual one-peak velocity distribution splits into two, asymmetrically positioned.  A surgical repair of the usual hydrodynamic description is presented that provides an account for, and an understanding of, the situation.   
\end{abstract}

\pacs{45.70.-n, 51.30.+i, 51.10.+y}
\maketitle

Loudspeakers operating in air and vibrating walls exciting granular gases are qualitatively different devices. The former perturbs  the equilibrium state of air slightly,  the latter provides the largest velocity in the system, and  is the very reason why the grains possess kinetic energy, why an associated temperature exists. In air at equilibrium, the velocity distribution $f(v)$ is a peak of width $\sqrt{T}$, centrally located. (We set $m$, the particle mass, to 1 throughout.) A loudspeaker displaces the peak periodically,  oscillating it by a velocity much smaller than its width, $\langle v\rangle\ll\sqrt{T}$. A vibrating wall hits the grains, injecting them into the gas. After numerous collisions absorbing much of the initial energy, the grains return to the wall's vicinity, to be hit again. Stationarily, when the energy injection equals dissipation, $f(v)$ does not change with time. It then has, close to the boundary, two peaks: One for the (+) particles leaving the wall swiftly, the other for the (--) particles returning rather more slowly. Away from the wall, the two peaks merge into one, because collisions equalize them. Yet this circumstance is not a mere boundary problem, as the inefficiency of energy transfer limits the size of a granular gas maintained by vibrating walls: If the system is too large, the center cools, forms clusters, and is no longer gaseous. If this is to be avoided, the two-peak structure extends fairly deep into the gas. That this circumstance is relevant for coming to terms with all aspects of granular gases has been emphasized especially by Evesque, see~\cite{evesque}.   

The data from \textbullet~2D simulations by Herbst et al~\cite{herbst2004}, who employ boundary conditions realistically modeling vibrating walls, and from \textbullet~ micro-gravity experiments obtained in an Novespace Airbus (Campaign 2006) in a  2D vibro-fluidized granular system~\cite{microgravity}, show two intriguing results that cry out to be understood. First, instead of being characterized by a pressure as any gas, the system, seemingly emulating a solid, develops an anisotropic stress -- such that $\sigma_{xx}$ normal to the pair of vibrating walls is strictly constant, while the tangential component $\sigma_{yy}(x)$ depends on $x$ -- though force balance is satisfied~\cite{herbst2004}. ($\sigma_{xx}$ is constant except in a genuine boundary layer of 1.5 grain diameters.) Second, the temperatures $T_x$ and $T_y$, for the velocity distribution along $\hat x$ and $\hat y$, are different. The reason is probably that it takes many lossy collisions before the injected momentum along $\hat x$ gets fed into the kinetic energy along $\hat y$, or $\hat z$. But the questions remain where the solid-like behavior comes from, and how to establish an equation of state that separately relates $\sigma_{xx}$ and  $\sigma_{yy}(x)$ to the  temperature and density fields, reproducing especially  $\sigma_{xx}$'s highly accurate constancy.

Conventional hydrodynamics employs as variables the densities of mass $\rho$ and momentum $\rho v$, both conserved, in addition to $T$, a measure for the averaged kinetic energy per particle in dilute systems, and more generally, for the averaged total energy in microscopic degrees of freedom. The pressure is  given as $P=\rho T$ in an ideal gas, and more generally as  $p=$ $(\rho\frac{\partial}{\partial\rho}-1)(w-Ts)$, where $w,s$ are respectively the energy and entropy density. 
Granular hydrodynamics~\cite{conv simulation,conv simulation2,conv simulation3}, as first proposed by Haff~\cite{haff}, has the same variables, though $T$ relaxes, going to zero if the grains' kinetic energy is not replenished. 

To account for a two-peak fluid, it seems obvious that one should employ as variables two different sets of $T,\rho,\rho v$, for the (+) and (--) particles, along with formulas for the pressures, $P_+$ and $P_-$, see~\cite{evesque2}. In addition,  one would need two additional sets of  $T,\rho,\rho v$ for $\hat y$ and $\hat z$. We are afraid this opens a Pandora box of state variables,  neither conserved nor truly independent, rendering the resultant theory arbitrary and unwieldy.  One should instead, we believe, retain the conserved variables $\rho$ and $\rho v$, also  $T$ -- as the average width of all peaks in the system~\cite{note}. 
It should then suffice, for a minimal, surgical modification, to introduce two additional variables. 

The first  is the distance between the peaks, which is the crucial second length of  a two-peak distribution, much more relevant than the difference between the two widths. (In a single-peak distribution, the width is the only length scale. But there are many more in a two-peak one, and providing the two widths alone is not sufficient to characterize the distribution.) Being a velocity difference, the new variable $\Delta_{i}$  is odd under time reversal and a vector. In our case, only $\Delta_{x}\not=0$.  

Second is the difference between $T_x$ and $T_y$, between the average width along $\hat x$ and the only width along $\hat y$. More generally, we have $\delta T_i\equiv T_i-T$, $i=x,y,z$, with $\sum_i\delta T_i=0$. These are -- similar to the order parameter of nematic liquid crystals~\cite{deGennes} --  the diagonal elements of a symmetric, traceless tensor. (They do not form a vector, because $T_x$ does not distinguish between $\hat x$ and $-\hat x$.) So a tensor, even under time reversal, needs to be added. Instead of $\delta T_{ij}$, however, we employ $t_{ij}$, the deviation of the granular temperature $T_g$ as considered below, see Eqs~(\ref{eq2},\ref{eq5}). In our case, only $t_{xx}=-t_{yy}\not=0$. 

Granular solid hydrodynamics (GSH)~\cite{granR2,granL3} was derived employing the hydrodynamic procedure. Relying on general principles valid irrespective how dense or rarefied  the system is, it leads to equations that include collisions and enduring contacts, are valid both in the dense, elasto-plastic limit, and the  rarefied one. For instance, GSH is capable of accounting for the relaxation of the temperature until it is zero. 

Although the present system deviates from rarefied gas in the opposite direction, towards ballisticity, the hydrodynamic procedure still works, if we add variables that characterize the deviation from local equilibrium. The reason is the hierarchy of equilibria: Although the two peaks or three widths are not in equilibrium with one another,  the elements within each are well thermalized. 

In deriving GSH, a granular heat $w$  is introduced  -- with $s_g$ the granular entropy, and $T_g\equiv\partial w/\partial s_g$ the associated temperature. It quantifies the energy contained in the mesoscopic, intergranular degrees of freedom, especially the strongly fluctuating part of the grains' kinetic and elastic energy. Expanding $w$ in $s_g$, we have
\begin{equation}
\label{eq2}
w=s_g^2/(2b\rho)=b\rho T_g^2/2,\quad b\sim(\rho_{cp}-\rho)^{a_1},
\end{equation}
with  $a_1=$ const. The lowest order term is quadratic because equilibrium, or minimal energy $w=0$, is given for $s_g=b\rho T_g=0$. (This is quite the same idea as with any Ginzburg-Landau energy functional, just without the fourth order term, or  a phase transition.)  
The density dependence of $b(\rho)$, with $\rho_{cp}$ the random close density, is chosen such that the associated pressure~\cite{granR2,granR3}, $P\equiv(\rho\frac{\partial}{\partial\rho}-1)(w-T_gs_g)$ $=-\frac12\rho^2T_g^2\partial b/\partial\rho$, given as
\begin{equation}\label{eq3}
P=\frac{a_1\rho\, w}{\rho_{cp}-\rho}= \frac{\frac12a_1\rho^2b\,T_g^2}{\rho_{cp}-\rho},
\end{equation}
is appropriate for all densities, see eg.~\cite{Bocquet}.

For a rarefied system, in which the elastic contribution to the energy is negligible, we may identifying the energy $w= \frac12b\rho T_g^2$ with the kinetic energy per unit volume $\rho T$ (for a 2D system), implying  $bT_g^2\sim 2T$, especially in the above expression for the pressure. Clearly, taking  $T_g\sim\sqrt  T$,  Haff's granular hydrodynamics is retrieved.

If the temperature is maintained by vibrating walls, we need (as discussed above) $\Delta_i$ and $t_{ij}$ as additional variables. They also contribute to the energy which, in an expansion in all three variables,  becomes 
\begin{eqnarray}\label{eq4}
w=(b\rho T_g^2+c\rho\Delta_i^2 +e\rho t_{ij}^2)/2.
\end{eqnarray}
These variables relax, specifically because they possess energy that may be redistributed among microscopic,  inner-granular degrees of freedom (such as phonons). The energy being quadratic, the relaxation stops when the variables are zero, and the energy vanishes. 

Taking $c,e=$ const, independent of the density, the pressure is not changed by introducing the new variables, and remains as given in Eq~(\ref{eq3}). (Same as with $P\sim\partial b/\partial\rho$, additional pressure contributions would have resulted from  $\partial c/\partial\rho$ and  $\partial e/\partial\rho$.) We assume this for simplicity, as we are more interested in an anisotropic stress, less in modifying a given pressure. 

\begin{figure}[b]
\begin{center}
\includegraphics[scale=0.5]{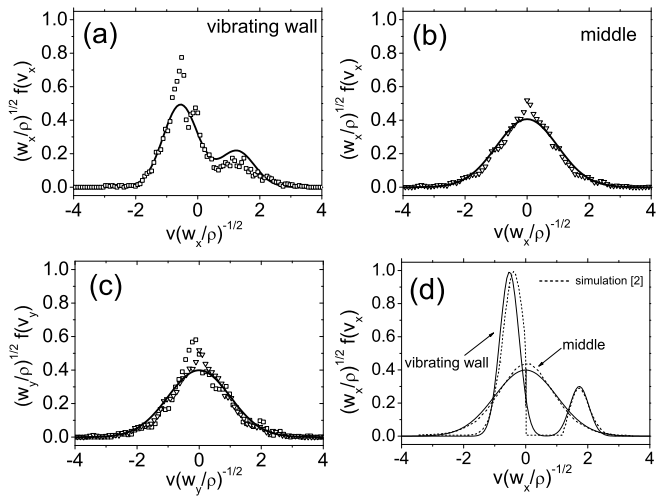}
\end{center}
\caption{Velocity distribution as simulated in~\cite{herbst2004}, measured in~\cite{microgravity}, and parameterized  by Eq~(\ref{EQ8}). Fig~(a) and (b) show $f(v_x)$, for the velocity perpendicular to the vibrating walls, with (a) showing  $f$ close to the wall, and (b) showing $f$ in the middle. Fig~(c) shows $f(v_y)$, while Fig~(d) again shows  $f(v_x)$. Symbols are measurements, dotted lines simulation, and full curves are Eq~(\protect\ref{EQ8}). For the two-peak distributions, we have $T_x=10$ and 1, $\protect\xi %
=-2.2$ and -1,37, $\protect\alpha =2.3$ and 3, for Fig~(a) and (d), respectively. }
\label{v-distr}
\end{figure}

Next we relate $\Delta _{x},t_{xx}$ to parameters of $f(v_{x})$ and $f(v_{y})$, the velocity
distributions, as the latter are independently measurable. Denoting the norm as $N\equiv \sqrt{\pi T_{x}}({1+\alpha })$ and $2k_B=1$, we take
\begin{equation}
f(v_{x})=\frac{1}{N}\left( {\alpha }\exp {\frac{(v_{x}-\xi )^{2}}{-T_{x}}}%
+\exp \frac{(v_{x}+\alpha \xi )^{2}}{-T_{x}}\right) ,  \label{EQ8}
\end{equation}
and $f(v_{y})=f(v_x\!\!\to\! v_y,T_{x}\!\!\to\!\! T_{y},\xi\! =\!0)$, with  $\langle v_{x}\rangle,\langle v_{y}\rangle =0$, see Fig~\ref{v-distr}. The
 energies along $\hat x,\hat y$ are then $w_{x}=\frac{1}{2}\rho T_{x}+\rho \alpha \xi ^{2}$ and  $w_{y}=\frac{1}{2}\rho T_{y}$, implying, first of all, $\alpha\!\!\to\! c,\,\xi\!\!\to\!\Delta_x$. 
 
Taking $T_x=\frac12(T+\delta T_{xx})$, $T_y=\frac12(T+\delta T_{yy})$, with $\delta T_{xx}+\delta T_{yy}=0$, $\xi=0$, we find that the energy, $w_{x}+w_{y}=\frac12\rho T$, does not depend on $\delta T_{xx}$. But taking
\begin{equation}\label{eq5}
T_{x}=b\left(T_g+t_{xx}\right) ^{2}\!\!/4, \,\, T_{y}=b\left(T_g-t_{xx}\right) ^{2}\!\!/4,
\end{equation}
we obtain  $w_{x}+w_{y}=\frac12\rho b (T_g^2+t_{xx}^2)$,  as in Eq~(\ref{eq4}).
This discrepancy may be surprising at first, but results from $\rho T$ being the energy of an ideal gas, or the kinetic energy of a rarefied gas, with no collisional contributions. Yet collisions are what equalize $T_x$ and $T_y$.  On the other hand, $w=\frac12\rho b T_g^2$ is the total energy of an interacting system. Increasing $T_g$ by  $t_{xx}$ in one population, and decreasing it by  $t_{xx}$ in another, must lead to an energy increase, as $t_{xx}$ would not relax otherwise. Still, we should not take the equality of $b$ and $e$ seriously, as it hinges on the precarious assumption that $w_{x}, w_{y}$ remain sensible quantities in an interacting system. In contrast, expanding $w$ in $t_{ij}$  leading to Eq~(\ref{eq4}) is a generally valid approach. 

Having specified the additional variables and their contributions to the energy, we may employ the same hydrodynamic procedure as used for GSH~\cite{granR2,granL3}, to set up their equations of motion.  The generalized hydrodynamics consists of continuity equations for momentum, $\partial_t(\rho v_i)+\nabla_j(\sigma_{ij}+\rho v_iv_j)=0$,  and mass,
$\partial_t\rho+\nabla_i(\rho v_i)=0$, in addition to the balance equations,  
\begin{eqnarray}\label{EQ3}
T_g[\partial_ts_g+\nabla_i(s_gv_i-\kappa_g\nabla_iT_g)]&=&\eta_gv^*_{ij}v^*_{ij}
-\gamma_gT_g^2,
\\\label{EQ4}
\partial_t\Delta_i+\nabla_j(\Delta_i v_j -\kappa_\Delta\nabla_j\Delta_i)&=& \alpha \Delta_jv^*_{ij}-\gamma_\Delta \Delta_i,
\\\label{EQ41}
\partial_t t_{ij}+\nabla_k(t_{ij} v_k -\kappa_t\nabla_k t_{ij})&=& \beta v^*_{ij}-\gamma_t t_{ij},
\\\label{EQ5}
\text{with}\,\,\quad\sigma_{ij}=P\delta_{ij}-\eta v^*_{ij}-c\rho&\alpha&\Delta_i\Delta_j-e\rho\,\beta\, t_{ij}.
\end{eqnarray}
The first equation is the same as in {\sc gsh}. It sports a convective, a diffusive ($\sim\kappa_g$) and a relaxative term ($\sim\gamma_g$), in addition to viscous heating, with  $\eta_g$ the viscosity, and $v^*_{ij}$ the shear rate -- $v^*_{ij}$ being the traceless part of $v_{ij}\equiv$ $\frac12(\nabla_iv_j+\nabla_jv_i)$.
Employing  $T_g\sim\sqrt{T}$, one sees that Eq~(\ref{EQ3}) is the same as Haff's energy balance~\cite{luding2009}. 
 Eqs~(\ref{EQ4},\ref{EQ41}) are new, but quite similar to~(\ref{EQ3}). They also each sport a convective, diffusive, and relaxative term. Instead of viscous heating, however, there is  a linear, off-diagonal Onsager term:  with $v_{ij}^*$ as the thermodynamic force, $\Delta_i$ as the preferred direction, and $\alpha$ an Onsager coefficient in Eq~(\ref{EQ4});  with  $v_{ij}^*$ as the force, no preferred direction, and $\beta$ another Onsager coefficient in  Eq~(\ref{EQ41}). 
 
The stress $\sigma_{ij}$ consists of pressure, viscous stress (with bulk viscosity neglected), and the two counter Onsager terms. The signs of $\alpha,\beta$ in the three equations obey Onsager reciprocity relation; and because both $\Delta_i\Delta_j v^*_{ij}$ and $v^*_{ij}t_{ij}$ are  odd under time inversion,  their respective contribution to the production of true  entropy (not displayed) vanish. There is no constraint on the sign or magnitude of $\alpha,\beta$; both are functions of the density.

\begin{figure}[b]
\begin{center}
\includegraphics[scale=.4]{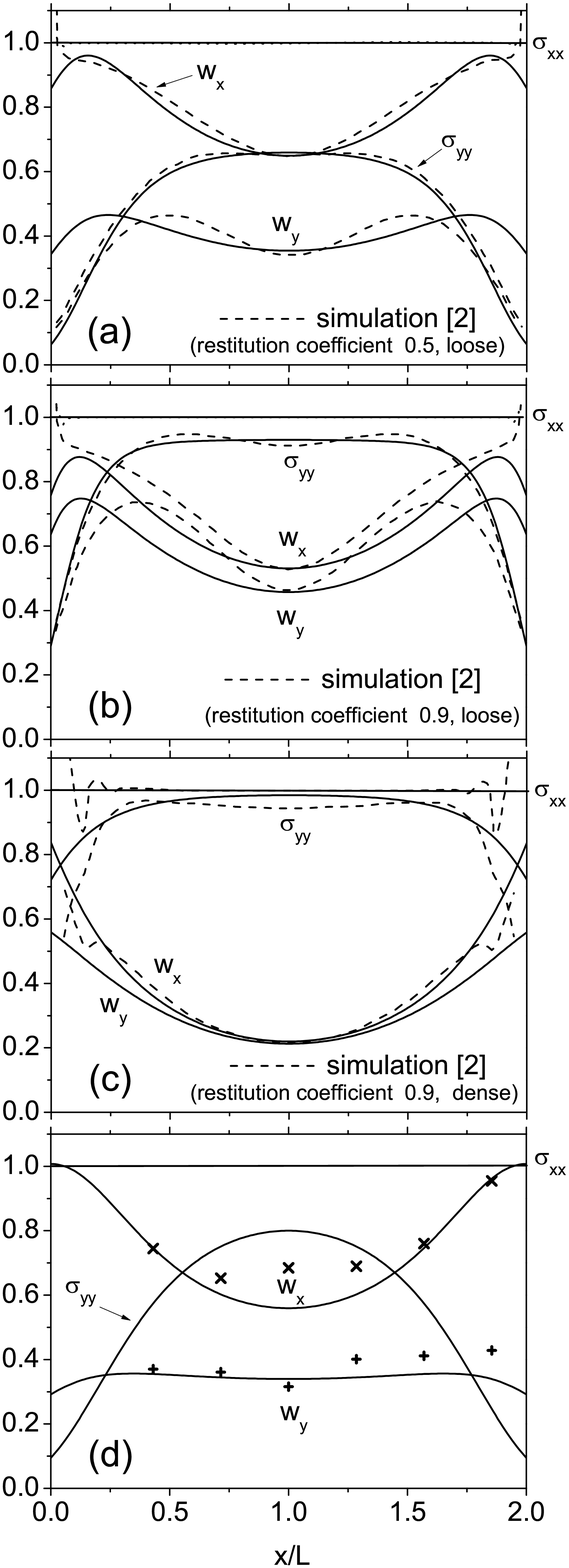}
\end{center}
\vspace{-.5cm}
\caption{ Variation of stress and kinetic energy along $\protect\hat{x}$. We employ $\sigma _{xx}=$ const as the unit of stress and energy density. Full curves are hydrodynamic results, taking $w_{x} =\rho bT_{g}^{2}/4+ \rho et_{xx}^{2}/2+ \rho c\Delta _{x}^{2}/2$, and
$w_{y} =\rho bT_{g}^{2}/4 +\rho et_{yy}^{2}/2$, with $b_{0},c,e=1$, and $\ell _{g}=1,1.1,2,1$, $%
\ell _{t}=0.8,0.9,1,0.5$, $\ell _{\Delta }=0.18,0.28,0.55,0.3$, $%
a_{1}=1.38,1.2,0.33,0.9$, $\rho _{cp}=0.7,0.7,0.7,0.8745$, $-\alpha
=13,1.49,0.772,0.5$, $-\beta =3.68,2.14,2.18,3.5$ for (a), (b), (c), (d),
respectively. Symbols are from
micro-gravity measurements of~\cite{microgravity}, and dotted lines from simulations of~\cite{herbst2004}.}
\label{profile}
\end{figure}

\begin{figure}[t]
\begin{center}
\includegraphics[scale=.4]{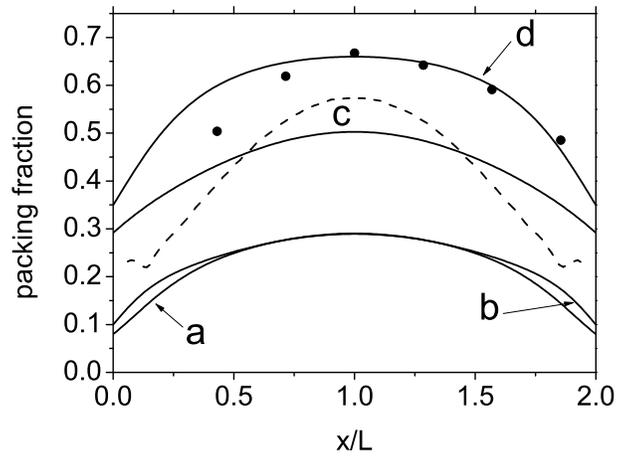}
\end{center}
\vspace{-.5cm}
\caption{Variation of the packing fraction for the four cases in Fig.2.  Symbols are from
micro-gravity measurements of~\cite{microgravity}, and dotted lines from simulations of~\cite{herbst2004}.}
\label{profile2}
\end{figure}

To solve Eqs~(\ref{EQ3},\ref{EQ4},\ref{EQ41},\ref{EQ5}), we first note that for the above discussed Herbst geometry, setting $v_i,v_{ ij}\equiv0$,  and assuming dependence only along $\hat x$, we have  $\sigma_{xy}=0$, 
\begin{equation}\label{EQ6}
\sigma_{xx}=P-c\rho\alpha\Delta_x^2-e\rho\beta t_{xx},\quad \sigma_{yy}=P(x)+e\rho\beta t_{xx}.
\end{equation}

Force balance $\nabla_j\sigma_{ij}=0$ requires $\sigma_{xx}=$ const, but leaves $\sigma_{yy}$ undetermined. Denoting  $\ell_g^2\equiv{\frac{\kappa_g}{2\gamma_g}}$, 
$\ell_\Delta^2\equiv{\frac{\kappa_\Delta}{2\gamma_\Delta}}$, 
$\ell_t^2\equiv{\frac{\kappa_t}{2\gamma_t}}$, with $2L$ the distance between 
the two vibrating walls,  and employing the boundary conditions:  $T_g=T_0$, $\Delta_x=\Delta_0$,  $t_{xx}=t_0$ at $x=0$, and  $T_g=T_0$, $\Delta_x=-\Delta_0$,  $t_{xx}=t_0$ at $x=2L$, the stationary solution, for $\partial_ts_g,\partial_t\Delta_i,\partial_tt_{ij}=0$, is
\begin{equation}
\!\frac{T_g}{T_0}=\frac{\cosh\frac{x-L}{\ell_g}}{\cosh\frac{-L}{\ell_g}}, 
\frac{t_{xx}}{t_0}=\frac{\cosh\frac{x-L}{\ell_t}}{\cosh\frac{-L}{\ell_t}}, 
\frac{\Delta_x}{\Delta_0}=\frac{\sinh\frac{x-L}{\ell_\Delta}}{\sinh\frac{-L}{\ell_\Delta}}.
\end{equation}
Note we have taken all transport coefficients, generally functions of $\rho, T_g, \Delta_i$, as constant. (Although $\kappa_g,\gamma_g\sim T_g$, see~\cite{granR2,luding2009}, this does not change the solution if included, since the equation contains only the ratio  $\ell_g^2\equiv\frac{\kappa_g}{2\gamma_g}$.) Searching for an understanding at present, we are unabashedly qualitative. Given the scarcity of experimental and simulation data, there is too much arbitrariness for more quantitative considerations. For a comparison of theory, experiment, and simulation, see Fig~\ref{profile},\ref{profile2}.

Summary: The physics of a granular gas has some idiosyncratic features, if its temperature is maintained by vibrating walls.  These are most notably first a two-peak distribution for the velocity perpendicular to the vibrating walls, second a much narrower peak width for the velocity along the walls, and third, as a result of the above two, a solid-like anisotropic stress.  We modified the hydrodynamic theory by introducing two additional variables: $\Delta_x$ for the distance between the two peaks, and $t_{xx}=-t_{yy}$ for the difference between the widths, where the former is part of a vector, and the latter part of a symmetric, traceless tensor. Similarly to the granular temperature $T_g$, both variables characterize the velocity distribution of an inelastic gas. All three diffuse and relax, displaying a similar macroscopic behavior. While the scalar $T_g$ gives rise to an hydrodynamic pressure, the vector $\Delta_i$ and the tensor $t_{ij}$ contribute to an anisotropic stress. Even under strongly simplifying assumptions, the calculated stress displays remarkable resemblance to that of simulations~\cite{herbst2004} and microgravity  experiments~\cite{microgravity}. We conclude that for this case,  the basic approach of a hydrodynamic description remains valid.

\acknowledgements
{\em Project supported by  National Natural Science Foundation of China (Grant
11034010) and the Special Fund for Earthquake Research of China (Grant 201208011).}

\begin{thebibliography}{99}
\bibitem{evesque}
P.~Evesque.
\newblock {\em Poudres \& Grains}, 18(1):1--19, 2010.

\bibitem{herbst2004}
O.~Herbst, P.~M\"uller, M.~Otto, and A.~Zippelius.
\newblock {\em Phys. Rev. E.}, 70:051313, 2004.
O.~Herbst, P.~M\"uller, and A.~Zippelius.
\newblock {\em Phys.Rev.E}, 72:041303, 2005.

\bibitem{microgravity} {Y.P. Chen}, P. Evesque,  {M.Y. Hou}, C. Lecoutre, F. Palencia, Y. Garrabos. 
J. Phys.: Conf. Ser.{\bf 327}, 012033 (2011). {Y.P. Chen}, P. Evesque,  {M.Y. Hou}. Chin. Phys. Lett.  {\bf 29}-7, 074501 (2012).

\bibitem{conv simulation}T. Poschell, S. Luding, 
Granular Gases, Lectures Notes in Physics 564, (Springer, Berlin, 2001);
Granular Gas Dynamics, Lectures Notes in Physics 624, edited by T. Poschel and N. V. Brilliantov, (Springer, Berlin, 2003); 

\bibitem{conv simulation2}S.Luding, R.Cafiero, H.J. Herrmann. 
Granular Gas Dynamics, Lectures Notes in Physics 624, edited by T. Poschel and N. V. Brilliantov, (Springer, Berlin, 2003); 

\bibitem{conv simulation3} A. Barrat, E. Trizac and M.H. Ernst, 
J. Phys. C (2005); 

\bibitem{haff}
P.~K. Haff.
\newblock {\em Journal of Fluid Mechanics Digital Archive}, 134(-1):401--430,
  1983.

\bibitem{evesque2} P. Evesque, 
{\em Poudres \& Grains} 20:1--26, 2012. 

\bibitem{note} For two groups of particles that do not interact, we have two separately conserved momenta, $\partial_t(\rho^+ v^+)+\nabla_j\sigma_{ij}^+=A$, $\partial_t(\rho^- v^-)+\nabla_j\sigma_{ij}^-=-A$, with $A=0$. If they interact weakly,  $A\not=0$ accounts for the momentum transfer. For strong interaction, there is only one conserved momentum, $\partial_t(\rho v)+\nabla_j\sigma_{ij}=0$, with $\rho v=\rho^+ v^++\rho^- v^-$. And it is not useful to separate $\sigma_{ij}^+$ or $\sigma_{ij}^-$ from $\sigma_{ij}$. 

\bibitem{deGennes}
P.G. de~Gennes and J.~Prost.
\newblock {\em The Physics of Liquid Crystals}.
\newblock Clarendon Press, Oxford, 1993.



\bibitem{granR2}
Y.M.~Jiang and M.~Liu.
\newblock {\em Granular Matter}, 11:139, 2009;
\bibitem{granR3}
Y.M.~Jiang and M.~Liu.
\newblock In D.~Kolymbas, G.~Viggiani, editors, {\em Mechanics of Natural
  Solids}, pages 27--46. Springer, 2009.
G.~Gudehus, Y.M. Jiang, and M.~Liu.
\newblock {\em Granular Matter}, 1304:319--340, 2011.

\bibitem{granL3}
Y.M.Jiang, M.~Liu.
\newblock {\em Phys.Rev.Lett.}, 99(10):105501, 2007.



\bibitem{Bocquet}
L.~Bocquet, W.~Losert, D.~Schalk, T.~C. Lubensky, and J.~P. Gollub.
\newblock {\em Phys. Rev. E}, 65(1):011307, Dec 2001.


\bibitem{luding2009}
Stefan Luding.
\newblock {\em Nonlinearity}, 22:101--146, 2009.

\end{thebibliography}

\end{document}